\documentclass[conference]{IEEEtran}
\IEEEoverridecommandlockouts
\usepackage{cite}
\usepackage{amsmath,amssymb,amsfonts}
\usepackage{algorithmic}
\usepackage{graphicx}
\usepackage{textcomp}
\usepackage{xcolor}
\usepackage{url}
\usepackage{hyperref}
\def\BibTeX{{\rm B\kern-.05em{\sc i\kern-.025em b}\kern-.08em
    T\kern-.1667em\lower.7ex\hbox{E}\kern-.125emX}}
\begin{document}

\title{BCDNet: A Fast Residual Neural Network For Invasive Ductal Carcinoma Detection
\thanks{\IEEEauthorrefmark{3}the corresponding author}
}

% \author{
% \IEEEauthorblockN{Yujia Lin}
% \IEEEauthorblockA{\textit{Glasgow College} \\
% \textit{University of Electronic Science and Technology of China} \\
% Chengdu, China \\
% linyujia@std.uestc.edu.cn}
% \and
% \IEEEauthorblockN{Aiwei Lian}
% \IEEEauthorblockA{\textit{Glasgow College} \\
% \textit{University of Electronic Science and Technology of China} \\
% Chengdu, China \\
% 2023190505038@std.uestc.edu.cn}
% \and
% \IEEEauthorblockN{Minyu Liao}
% \IEEEauthorblockA{\textit{Glasgow College} \\
% \textit{University of Electronic Science and Technology of China} \\
% Chengdu, China \\
% 2023190505019@std.uestc.edu.cn}
% \and
% \IEEEauthorblockN{Yipeng Liu \IEEEauthorrefmark{1}}
% \IEEEauthorblockA{\textit{School of Information and Conmmunication Engineering} \\
% \textit{University of Electronic Science and Technology of China} \\
% Chengdu, China \\
% yipengliu@uestc.edu.cn}
% }

\author{
\IEEEauthorblockN{
Yujia Lin \IEEEauthorrefmark{1}\textsuperscript{1}, 
Aiwei Lian \IEEEauthorrefmark{1}\textsuperscript{2}, 
Mingyu Liao \IEEEauthorrefmark{1}\textsuperscript{3}, 
and Shuangjie Yuan \IEEEauthorrefmark{2}\textsuperscript{1}}
\IEEEauthorblockA{\IEEEauthorrefmark{1}Glasgow College, University of Electronic Science and Technology of China, Chengdu, China \\
\textsuperscript{1}linyujia@std.uestc.edu.cn, 
\textsuperscript{2}2023190505038@std.uestc.edu.cn, 
\textsuperscript{3}2023190505019@std.uestc.edu.cn}
\IEEEauthorblockA{\IEEEauthorrefmark{2}School of Automation Engineering, University of Electronic Science and Technology of China, Chengdu, China \\
\textsuperscript{1}sj.yuan.uestc@gmail.com}
}

\maketitle

\begin{abstract}
It is of great significance to diagnose Invasive Ductal Carcinoma (IDC) in early stage, which is the most common subtype of breast cancer. Although the powerful models in the Computer-Aided Diagnosis (CAD) systems provide promising results, it is still difficult to integrate them into other medical devices or use them without sufficient computation resource. In this paper, we propose BCDNet, which firstly upsamples the input image by the residual block and use smaller convolutional block and a special MLP to learn features. BCDNet is proofed to effectively detect IDC in histopathological RGB images with an average accuracy of 91.6\% and reduce training consumption effectively compared to ResNet 50 and ViT-B-16.
\end{abstract}


\section{Introduction}
In 2020, there were 2.26 million new breast cancer cases and 0.684 million deaths reported globally, making breast cancer the most frequently diagnosed cancer and the fourth leading cause of cancer-related deaths among 36 cancer types \cite{kwong_survey_2022}. The incidence of breast cancer has been rising by 0.5\% annually from 2010 to 2019 \cite{noauthor_breast_nodate}. Invasive Ductal Carcinoma (IDC) is the predominant subtype of breast cancer, accounting for 80\% of all cases \cite{wang_early_2017}.

Despite the availability of various diagnostic methods such as mammography, digital breast tomosynthesis, breast ultrasound, and magnetic resonance imaging, accurately and rapidly diagnosing the large scale of IDC cases remains a significant challenge for radiologists and other medical staff. To assist pathologists, Computer-Aided Diagnosis (CAD) systems have been developed as complementary tools to conventional diagnostic methods \cite{rangayyan_review_2007,tang_computer-aided_2009}. These systems analyze medical images or other data and provide diagnostic results to pathologists, specifically aiding in the diagnosis of IDC by saving time while maintaining accuracy.

Deep Learning has greatly advanced Computer-Aided Diagnosis (CAD) by now. In Medical Image Processing, CNN-based methods are widely adopted for breast cancer detection due to CNNs' ability to automatically learn features, eliminating the need for hand-crafted features. In detecting IDC, histopathological image features are influenced by factors such as geographical region, ethnicity, and patient age \cite{noauthor_breast_nodate}, \cite{wang_early_2017}. However, public datasets often fail to fully capture the variety in these images, leading to inaccurate predictions when generalized models are used. Therefore, fine-tuning models on real-world datasets is crucial for improving detection accuracy. At the same time, the demand of integrated and automatic diagnosis system is increasing, requiring the deep learning models to be combined with other software and hardware components to provide a seamless and efficient solution.

Therefore, many existing approaches that use large models is not suitable for this task. A powerful and expensive computer is required to train and deploy such models, which is not always feasible in real-world scenarios. This limitation prevents their practical application in scenarios where fast diagnosis is needed or computing resources are constrained. To address these challenges, we developed BCDNet—a model designed to be both computationally efficient and easy to train. We achieve this by designing the architecture again. In our BCDNet, the input image is firstly upsampled to make it easier to extract features. Then the features are learned by a small convolutional block and a improved MLP. This allows for reducing the parameters and the deployment on edge devices with limited computation ability, enabling fast adaptation to new datasets and making it suitable for practical and real-time medical use. Our main contributions can be concluded as follows:
\begin{itemize}
    \item We propose a novel architecture BCDNet, which is efficient and suitable for small-scale datasets.
    \item We reassemble the widely used neural network architectures for Breast Cancer Detection to make BCDNet mission-critical.
    \item Our model still remains simple and lightweight with almost the same performance as other complex models.
\end{itemize}

\section{Related Work}

\subsection{Conventional Diagnosis Methods}
To diagnose breast cancer, four conventional methods are commonly employed: Mammography, Digital Breast Tomosynthesis (DBT), Breast Ultrasound, and Magnetic Resonance Imaging (MRI) \cite{kwong_survey_2022,wang_early_2017,mcdonald_clinical_2016,noauthor_sensors_nodate,noauthor_journal_nodate,noauthor_breast_nodate-1}. Mammography, an initial method for early-stage breast cancer detection using X-ray to obtain 2D images, is divided into screening mammography and Digital Mammography (DM). As imaging technology rapidly advances, new diagnostic opportunities emerge. DBT, more mature than DM, offers a novel way of observing breast cancer images by capturing 2D slices and synthesizing them into 3D images. This reduces tissue overlap and enhances diagnostic accuracy. Automatic Breast Ultrasound (ABUS) enables radiologists to diagnose breast cancer with greater accuracy through improved imaging frequency. Additionally, MRI diagnoses breast cancer using magnetic fields and radio waves to capture detailed images.

\subsection{Deep Learning-Based Methods}
In 2016, Wang et al. \cite{wang_deep_2016} introduced a segmentation method to localize potential cancer areas and used GoogLeNet as a backbone for classification. Subsequently, Chougrad et al. \cite{chougrad_deep_2018} proposed a method that enabled several state-of-the-art (SOTA) architectures to outperform traditional ones on the MIAS database, utilizing fine-tuning techniques to adapt ResNet50, VGG16, and Inception v3 for breast cancer detection.

While these methods show excellent performance in laboratory settings, many medical institutions, especially in developing countries, face a lack of computing resources and rich datasets, impeding the clinical application of SOTA models. To address these limitations, we propose BCDNet that is easy to train without rich computing resource.

\section{Method}

\begin{figure*}[!t]
    \centering
    \includegraphics[width=0.9\textwidth]{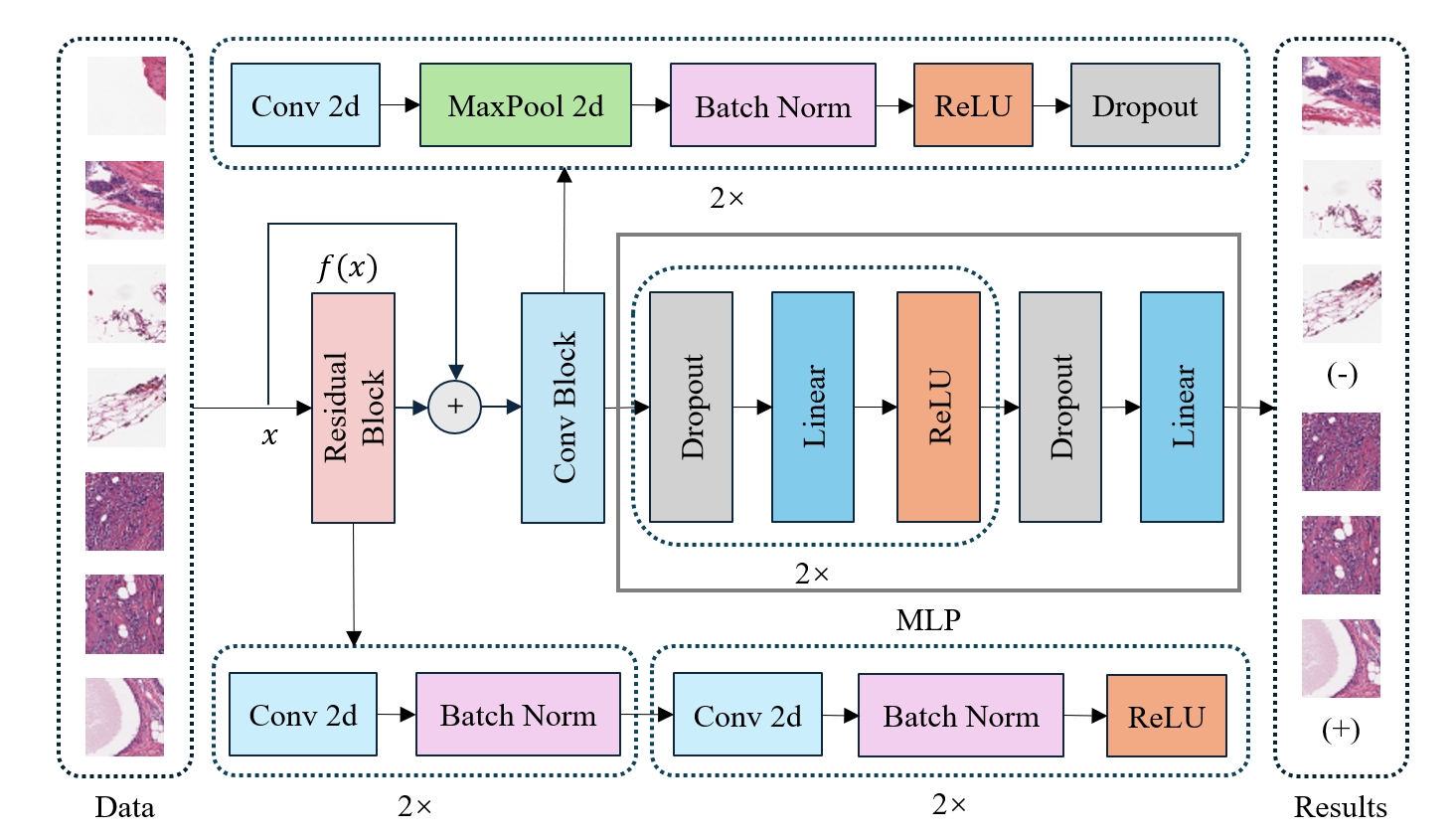}
    \caption{The architecture of BCDNet. We combine residual blocks, convolutional blocks, and MLPs to extract features from the input image. The detailed structure of each block is alsoshown in the figure. The blocks are generally composed of convolutional layers, pooling layers, activation layers, normalization layers, and dropout layers in different orders.}
    \label{fig1}
\end{figure*}

To improve the real-time performance and decrease the memory usage while remaining a high accuracy, we propose the architecture of BCDNet, which is composed of a residual block, a convolutional block and a special designed Multi Layer Perceptron (MLP). The residual block will initially upsample the input image so that we can decrease the depth and width of convolutional block to decrease the number of parameters. After the residual block, we use a smaller convolutional block and MLP with Dropout layer to further improve the speed while maintaining the accuracy. In the following sections, we will delve into the specific configurations and functionalities of each layer type within BCDNet.

\subsection{Residual Block}
The residual block is designed for increase the number of the channels of images to learn more necessary and complex features. Since the IDC histopathological images are basically the same, it is of great significance for the model to recognize the tiny differences between them. One common way is to increase the channel of the input images by applying the convolutional layers which will not change the size of the input image. The convolutional layer is the fundamental componet of the modern CNNs \cite{li_survey_2022,oshea_introduction_2015,albawi_understanding_2017} and can be represented as:
\begin{equation}
    Y(i,j) = \sum_{m=1}^{k_w} \sum_{n=1}^{k_h} X(i+m-1, j+n-1) \cdot K(m,n)
\end{equation}
where $X(i,j)$ and $Y(i,j)$ are the input and output matrix respectively; $K(m,n)$ is the kernel; $k_w$ and $k_h$ are the width and height of the kernel, which are equal in most situations.

In the residual block, Batch Normalization is used after the convolutional layers, enabling the model to learn more robust features and boost the speed of convergence \cite{ioffe_batch_2015}. Additionally, it can partly issues like exploding gradients. We simply denote Batch Normalization layer as:
\begin{equation}
    y_i = \frac{\gamma (x_i - \mu)}{\sigma} + \beta
\end{equation}
where $y_i$ is the normalized output, $\gamma$ and $\beta$ are learnable parameters, $\mu$ and $\sigma$ are the mean and standard deviation of the small batch of input, respectively.

To avoid vanishing gradient problem seen in sigmoid and tanh functions, we choose the Rectified Linear Unit (ReLU) as the activation function in the residual block \cite{glorot_deep_2011}. The function is defined as the following equation and can be seen in, equates values less than zero to zero, and values greater than or equal to zero to passed in value. 
\begin{equation}
    f(x) = max(0,x) = 
    \begin{cases}
        0, x \leq 0 \\
        x, x > 0
    \end{cases}
\end{equation}

To further improve the performance of BCDNet, we add a residual connection. In 2016, He et al. \cite{he_deep_2016}, proposed the Deep Residual Network that contains the residual connection, which is then become the state-of-the-art architecture for image classification. The connection reformulates the layers as learning residual functions with reference to the layer inputs, rather than directly learning unreferenced functions. Applying this shortcut in the neural network allows the model to learn more complex features and decrease the probability of degradation problems as network depth increases.

Mathematically, if $H(x)$ is the desired mapping, the residual block instead learns:
\begin{equation}
    H(x) = F(x) + x
\end{equation} 
This simple addition allows the network to pass information directly from earlier layers to later layers, thereby facilitating gradient flow during backpropagation and improving the ability to train very deep networks. The residual connections also enable better preservation of low-level features throughout the network, which is beneficial for accurate representation learning. Meanwhile, the original input is upsampled to match the dimensions of the output, which is achived by a $1 \times 1$ convolution.

\subsection{Convolutional Block}
By leveraging the residual block, the features in the input image are learned in different channels of the feature maps. However, the features are not extracted and the feature maps are still too large for practical using. Therefore, we build a convolutional block to extract the features from the input image and reduce the spatial dimensions of the feature maps. 

The convolutional block is composed of several convolutional layers, pooling layers, activation layers, and normalization layers. The pooling layers here are Max Pooling layers, which can be denoted as:
\begin{equation}
    M(i,j) = \max_{0 < p, q \leq k} F_{i+p,j+q}
\end{equation}
where $M$ is the down-sampled feature map, and $M_{(i,j)}$ is the maximum value within the $k \times k$ window centered at $(i,j)$ in the original feature map. The Max Pooling layer will preserve the maximum value within the window, which means it can extract the most outstanding features in the feature map. Moreover, the pooling layers can also reduce the spatial dimensions of the feature maps, which is beneficial for reducing the computational complexity and improving the model's efficiency.

In the convolutional block, we also add several Dropout layers to prevent overfitting and make the model more robust and lightweight \cite{gal_dropout_2016}. By randomly setting a fraction p of the neurons' activations to zero during the training, the network is encouraged to learn more robust features, thereby improving its generalization capabilities. The dropout process can be expressed as:
\begin{equation}
    h_i = 
    \begin{cases}
        0, \text{with probability } p \\
        \frac{x_i}{1-p}, \text{otherwise}
    \end{cases}
\end{equation}
Here, $p$ is the dropout rate, and the scaling factor $1/(1-p)$ ensures that the expected sum of the activations remains consistent during both training and inference.

\subsection{Multi-Layer Perceptron (MLP)}
Though we regard the Breast Cancer Detection as a binary classification task, we hope that other scholars and medical professionals can also use our model to perform more accurate classification about the certain subtype of Breast Cancer. Thus, we design a more complex MLP than this binary classification task needs. Based on the traditional MLP, we add dropout layers to prevent overfitting and make the model more robust and lightweight. ReLU is also used as the activation function in the MLP, which can make this MLP learn more non-linear relationships between the input features and the output. The structure of this MLP can be seen in Fig \ref{fig1}.

\section{Experiments}
To evaluate BCDNet's performance, we conducted a comprehensive experiment between BCDNet, ResNet50 and ViT-B-16, which are renowned for their robust performance in image classification tasks. ResNet50, a deep residual network, is celebrated for its pioneering use of residual learning to solve vanishing gradient problems \cite{he_deep_2016}, while ViT-B-16, a Vision Transformer, represents a shift towards transformer architectures in computer vision, demonstrating strong performance by leveraging self-attention mechanisms \cite{han_survey_2023}. 

Training was conducted on a server with 2 NVIDIA RTX 4090 GPUs, leveraging a cross-entropy loss function and the Adam optimizer. The learning rate was initially set to 0.005, with a StepLR scheduler to facilitate convergence. Our study is based on two open-source IDC image datasets, and we assess models on accuracy, training time, and memory consumption.

\begin{figure}[htbp]
    \centering
    \includegraphics[width=0.5\textwidth]{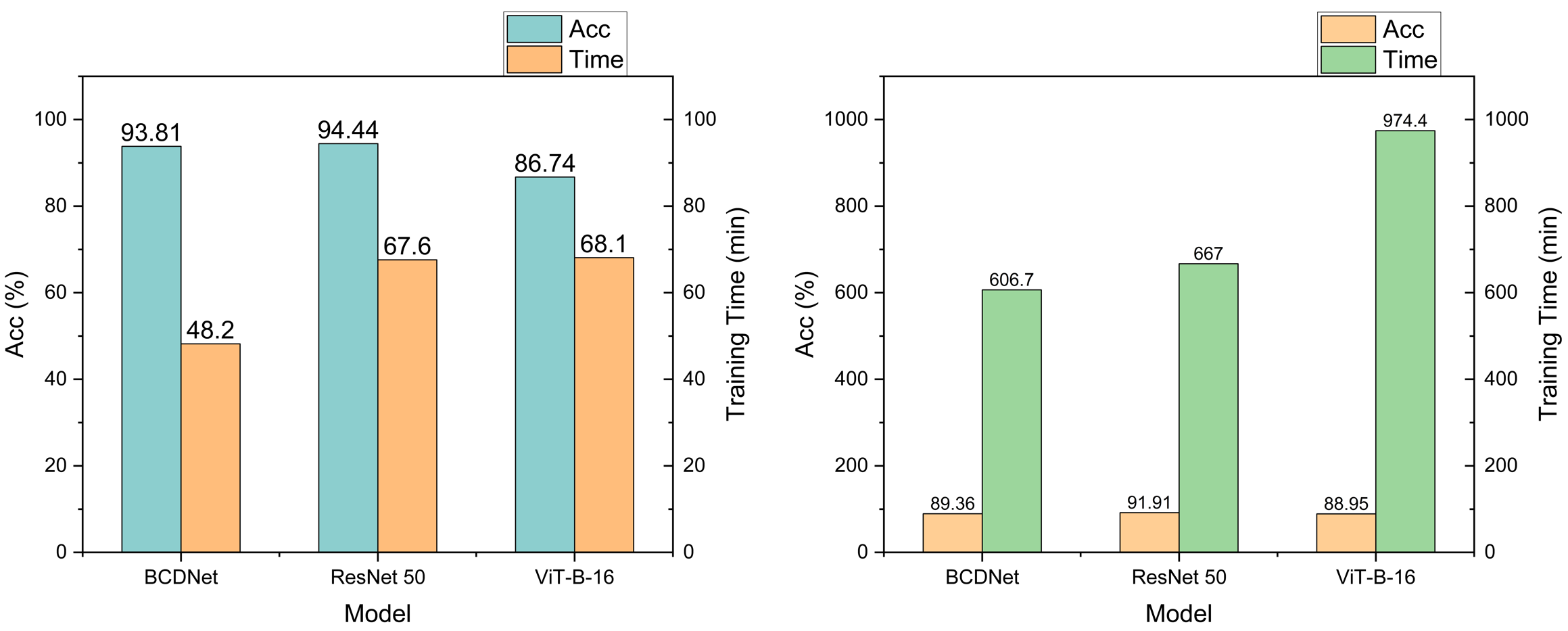}
    \caption{The accuracy and training time of BCDNet, ResNet 50 and ViT-B-16. \textbf{Left}: The results of the three models on BreaKHis v1 dataset. \textbf{Right}: The results of the three models on IDC regular dataset.}
    \label{fig2}
\end{figure}

\begin{figure}[htbp]
    \centering
    \includegraphics[width=0.5\textwidth]{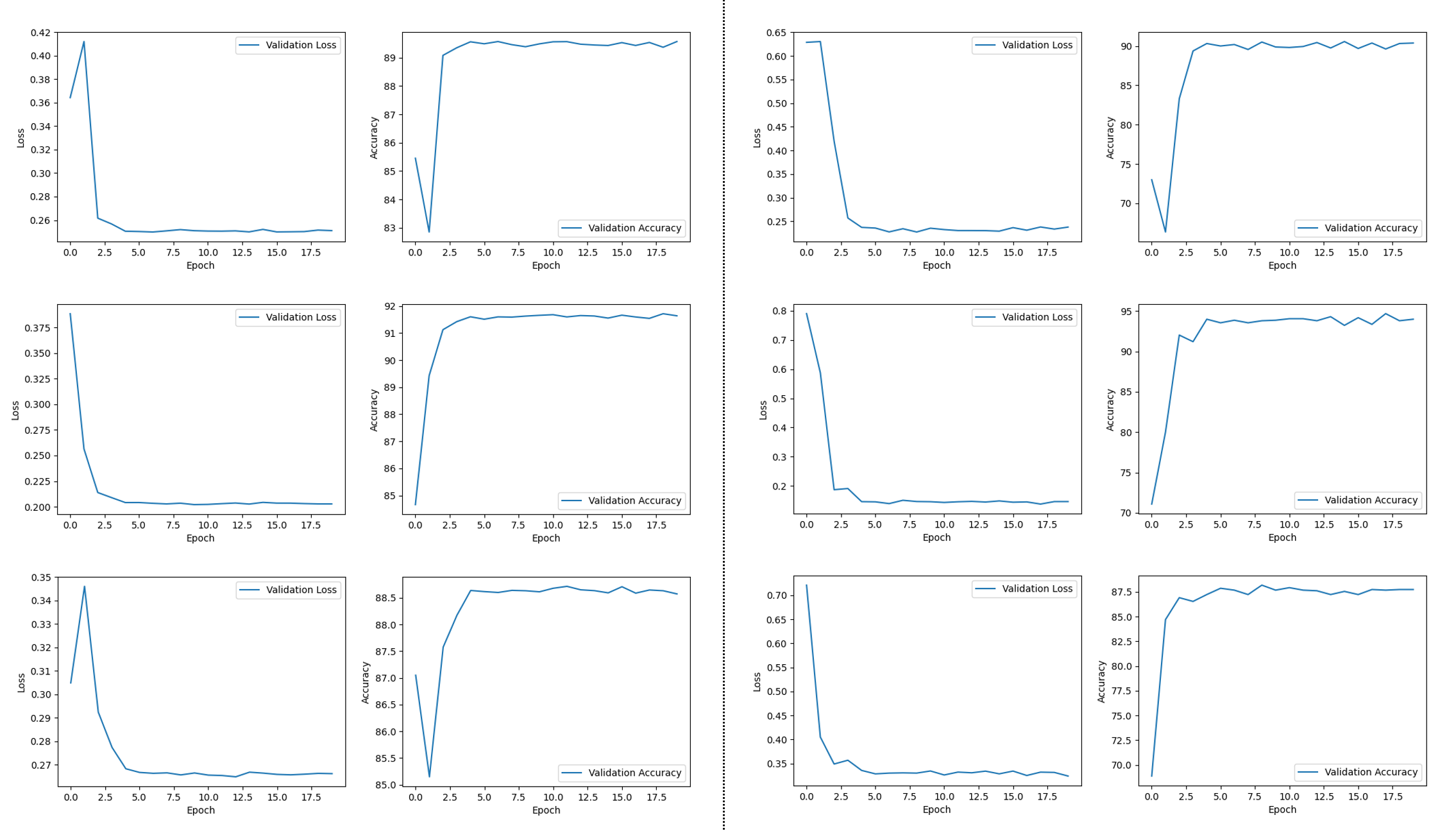}
    \caption{The curves of the accuracy and loss of BCDNet, ResNet 50 and ViT-B-16 (from the first row to the third row). \textbf{Left}: The curves of the three models on IDC regular dataset. \textbf{Right}: The curves of the three models on BreaKHis v1 dataset.}
    \label{fig3}
\end{figure}

\subsection{Dataset}
In this study, we firstly utilized an IDC image dataset from kaggle, which is termed as \textit{IDC regular} by us in this paper. IDC regular dataset comprises histopathological scans of breast cancer tissues in RGB format, annotated to indicate the presence or absence of IDC. Meanwhile, we also use \textit{BreaKHis v1} dataset, which is composed of multiple types of breast cancer histopathology images obtained under various imaging conditions, including different stains, magnifications, and imaging protocols. The two datasets are divided into training, validation and testing sets by a proportion of 7:2:1 respectively to enable the evaluation of our proposed model, BCDNet, against established benchmarks.

\subsection{Data Augmentation}
In our study, we first applied Normalization to every image to accelerate the training. Then, we used random horizontal and vertical flipping and random rotation to increase the diversity of the training set. Finally, all the images were resized to $224 \times 224$ pixels to ensure consistency in the input dimensions.

\subsection{Evalution Metrics}
We evaluated the models based on accuracy, training time, and GPU memory consumption. Training time and memory usage were monitored to highlight the efficiency of each model. Accuracy was assessed on an independent test set to ensure fair comparison.

\subsection{Results}
Based on Fig \ref{fig2}, Fig \ref{fig3} and Table \ref{tab1}, we can summarize that though BCDNet has a slightly lower accuracy than ResNet 50, it is more efficient in terms of training time and memory consumption even though we used pretrained models of ResNet and ViT in our experiments. BCDNet is more suitable for deployment on edge devices and can be quickly modified for new datasets. Moreover, the convergence speed of BCDNet is also higher, since its loss and accuracy both become stable quite early.

\begin{table}[htbp]
    \caption{The Memory Consumption of BCDNet, ResNet 50 and ViT-B-16.}
    \begin{center}
    \begin{tabular}{|c|c|c|}
    \hline
    \textbf{Model}&\multicolumn{2}{|c|}{\textbf{Datasets}} \\
    \cline{2-3} 
    \textbf{Types} & \textbf{IDC regular} & \textbf{\textit{BreaKHis v1}} \\
    \hline
    BCDNet & 14.8 GB & 14.5 GB \\
    \hline
    ResNet 50 & 16.7 GB & 16.6 GB \\
    \hline
    ViT-B-16 & 20.8 GB & 20.8 GB \\
    \hline
    \end{tabular}
    \label{tab1}
    \end{center}
\end{table}

\section{Conclusion}
In this paper, BCDNet has demonstrated several key advantages, including reduced training time and lower memory consumption, all while maintaining the level of accuracy. Basically, our method can save 12.5\% GPU memory than ResNet 50 and 28.8\% than ViT-B-16 under our training settings. In terms of time, it is averagely 1.40 times faster than the two baselines on BreaKHis v1 dataset, and 1.35 times faster on IDC regular dataset. The accuracy of BCDNet on two datasets are 93.8\% and 89.3\% respectively. From our experiments, we can conclude that our pipeline is efficient and effective for IDC detection, which means adding the number of channels and extract features using a shallower convolutional block can accelerate the training process and not decrease the accuracy heavily.

\bibliographystyle{IEEEtran}
\bibliography{ref.bib}

\end{document}